\documentclass{aa}  
\usepackage{amsmath}
\usepackage{graphicx}
\usepackage{txfonts}
\begin{document}
   \title{An extremely wide and very low-mass pair 
          with common proper motion\thanks{Based on
          observations with the ESO~3.6m/EFOSC2 at the European Southern
          Observatory, La Silla (ESO program 70.C-0568)}}

  \subtitle{Is it representative of a nearby halo stream?}

   \author{R.-D. Scholz
          \inst{1}
          \and
          N.V. Kharchenko
          \inst{1,2}
          \and
          N. Lodieu
          \inst{3}
          \and
          M.J. McCaughrean
          \inst{4}
          }

   \offprints{R.-D. Scholz}

   \institute{Astrophysikalisches Institut Potsdam,
              An der Sternwarte 16, D-14482 Potsdam, Germany\\
              \email{rdscholz@aip.de, nkharchenko@aip.de}
         \and
              Main Astronomical Observatory, 
              27 Akademika Zabolotnogo St., 03680  Kiev, Ukraine\\
              \email{nkhar@mao.kiev.ua}
         \and
              Instituto de Astrofisica de Canarias,
              C/ Via Lactea s/n, E-38205 La Laguna, Tenerife, Spain\\
              \email{nlodieu@iac.es}
         \and
              Astrophysics Group, School of Physics, University of Exeter,
              Stocker Road, Exeter EX4 4QL, UK\\
              \email{mjm@astro.ex.ac.uk}
             }

   \date{Received 18 March 2008; accepted ...}

 
  \abstract
   {}
   {We describe the discovery of an extremely wide pair of low-mass 
   stars with a common large proper motion and discuss their possible
   membership in a Galactic halo stream crossing the Solar neighbourhood.}
   {In a high proper motion survey of the southern sky we used 
   multi-epoch positions and photometry from the SuperCOSMOS Sky Surveys.
   New nearby ultracool dwarf and subdwarf candidates were selected 
   among the faint and red high proper motion objects, and subsequently 
   confirmed by low-resolution classification spectroscopy. The resulting  
   spectroscopic distance estimates, 
   approximate 
   radial velocity measurements and
   improved proper motions involving additional epochs from 
   the Two Micron All Sky Survey and from the DEep Near-Infrared Survey
   were used to compute Galactic space velocities.}
   {The late-type (M7) dwarf SSSPM\,J2003$-$4433 and the ultracool subdwarf
   SSSPM\,J1930$-$4311 (sdM7) sharing the same
   very large proper motion of about 860~mas/yr were found in the same sky
   region with an angular separation of about 6\degr. 
   From the comparison with other high proper motion catalogues we 
   have estimated the probability of a chance alignment of the two new large 
   proper motions to be less than 0.3\%.
   From the individually estimated spectroscopic distances 
   of about $38^{+10}_{-7}$~pc and $72^{+21}_{-16}$~pc,
   respectively for the M7 dwarf and the sdM7 subdwarf, 
   and in view of the accurate
   agreement in their large proper motions we assume a common distance of
   about 50~pc and a projected physical separation of about 5~pc. 
   The mean 
   heliocentric
   space velocity of the pair $(U,V,W)=(-232, -170, +74)$~km/s,
   based on 
   the correctness of the
   preliminary radial velocity
   measurement for only one of the components 
   and on the assumption of a common distance and velocity vector,
   is typical of the Galactic halo population.}
   {The large separation and the
   different metallicities of dwarfs and subdwarfs make a common formation
   scenario as a wide binary (later disrupted) improbable,
   although there remains some uncertainty in
   the spectroscopic classification scheme of ultracool dwarfs/subdwarfs 
   so that a dissolved binary origin cannot be fully ruled out yet. 
   It seems more likely 
   that this wide pair is part of an old halo stream.
   Higher-resolution
   spectroscopic observations are needed to measure accurate radial velocities
   of both components. Further, we suggest to check the M7 dwarf for an
   unresolved binary status, which would explain its shorter spectroscopic
   distance estimate, and to place both objects on a trigonometric parallax
   program.}

   \keywords{Stars: kinematics --
                Stars: low-mass, brown dwarfs --
                Galaxy: formation --
                Galaxy: halo --
                solar neighbourhood
               }

\titlerunning{An extremely wide and very low-mass pair with common proper motion}

\authorrunning{R.-D. Scholz et al.}

   \maketitle
%

\section{Introduction}

Common proper motion (CPM) stars are typically wide binary stars for 
which the measuring errors of the proper motion are much larger than 
the effect of orbital motion. Wide binaries of different masses 
are interesting for stellar evolution studies since their components, 
believed to have the same age and metallicity, have evolved without 
interaction and can be studied independently and compared to each other 
(e.g. Seifahrt et al.~\cite{seifahrt05}; Catal\'{a}n et al.~\cite{catalan08}).
Halo wide binaries provide important traces of the formation of the
Galaxy (Allen et al. \cite{allen07}) and can be used to constrain 
Galactic halo dark matter (Chanam{\'e} \& Gould \cite{chaname04}).

CPM stars are usually found among stars with
large proper motions where the compatibility of the proper motions is
easier to detect. The ratio between the angular separation and the 
proper motion can be used as a criterion to discriminate between
physical and optical pairs (Halbwachs \cite{halbwachs86}). But even 
for very large separations, a physical connection in terms of membership
in a cluster or stellar stream 
(or a dissolved binary status)
is obvious if the individual proper motions 
are much larger than their measuring errors and agree within these errors.

The LDS (Luyten Double Stars) catalogue of CPM stars 
(Luyten~\cite{luyten4087}) lists 20 pairs with angular separations larger
than 600~arcsec (maximum 9000~arcsec). Only four of these pairs 
have a proper motion above the lower limit of the Luyten Half Second 
(LHS; Luyten~\cite{luyten79a}) 
catalogue of 500~mas/yr. For LDS entries with smaller proper motions and 
relatively large separations there is a higher probability of random pairings 
(see e.g. LDS\,4041 and LDS\,4999 in Scholz, Meusinger \& 
Jahrei{\ss} \cite{scholz05}). The catalogue of wide binaries by Allen, 
Poveda \& Herrera (\cite{allen00}) includes a few very wide CPM pairs 
with angular separations up to about 800~arcsec and with a maximum expected 
semi-major axis
of about 125\,000~AU.
At least the primary components are typically bright Hipparcos stars
so that the projected physical separations given in that catalogue rely on
trigonometric parallax measurements. 

Extremely wide (angular separation $>$1\degr) binaries and multiple systems 
with pre-Hipparcos trigonometric parallaxes were presented 
in Gliese \& Jahrei{\ss} (\cite{gliese88}). In their list of 32 systems with
projected tangential separations exceeding 5000\,AU, there are 9 systems 
with angular separations between 1\degr\ and 16\degr, corresponding to
minimum physical separations of 10\,000\,AU up to more than a million AU
(their proper motions range between 0.2 and 3.8~arcsec/yr). Four of the 
latter were discussed by the authors as gravitationally unbound due 
to their large separations and differences in their parallaxes
or radial velocities. The Hipparcos parallaxes (ESA~\cite{esa97}) 
have shown that only 3 out of the above mentioned 9 systems,
Gl~881+879, Gl~803+799AB, and Gl~559AB+551, can be considered
as physical since their parallaxes agree to within 5\% and 
their projected linear separations are 54\,250, 47\,180, and 10\,380~AU,
respectively\footnote{Three other systems (Gl~331ABC+332AB, 
Gl~48+22AB, Gl~1255ABC+HD~199476) are questionable physical systems
with projected linear separations between 100\,000 and 350\,000~AU
and parallax differences of 10-15\%.}.
It is remarkable that for the first two of these systems, which are
located at distances of 8 and 10~pc, respectively, all proper
motion components do also agree to within about 10~mas/yr. This is
not the case for Gl~559AB+551 ($= \alpha$CenAB+Proxima), where the
distance of only 1.3~pc and the orbital motions result in differences
up to a few 100~mas/yr between the three components of this bound
system (Wertheimer \& Laughlin \cite{wertheimer06}).

%
%
%
%
%
%
%

All previously known extremely wide systems include at least one 
star with $>$0.5~$M_{\odot}$.
In this paper, we present the first extremely wide CPM
pair consisting of two very low-mass ($\approx$0.1~$M_{\odot}$) stars. 
In Section~\ref{sect_cpm} 
we describe the astrometric measurements and show how accurately 
the large proper motions of the two faint red stars separated by 
about 6\degr\ agree with each other. 
In Section~\ref{sect_spec} we present a rather surprisingly
different spectroscopic classification of the components of this 
extremely wide CPM pair, and in Section~\ref{sect_halo}
we discuss their possible membership in a Galactic halo stream. 


\section{Common proper motion}
\label{sect_cpm}

The two high proper motion (HPM) objects SSSPM\,J1930$-$4311 (hereafter 
SSSPM\,1930) and SSSPM\,J2003$-$4433 (hereafter SSSPM\,2003) were discovered 
in the course of a HPM survey (described in Lodieu et al.~\cite{lodieu05}) 
using multi-epoch SuperCOSMOS 
Sky Surveys (SSS) data (Hambly et al.~\cite{hambly01}). The 3rd search strategy 
outlined in Lodieu et al.~(\cite{lodieu05}; section 2) was applied, i.e only
$R$ and $I$ band data were used for the HPM detection. The proper motion
solution for both objects was later improved by including all available 
SSS data as well as the positions measured in the Two-Micron All Sky Survey 
(2MASS; Cutri et al.\ \cite{cutri03}), in the DEep Near-Infrared
Survey (DENIS; Epchtein et al.\ \cite{epchtein97}), using the third DENIS data 
release from 2005, and the positions of our targets on the acquisition
images taken with the ESO 3.6m telescope. The latter were measured by us
using the 2MASS positions of 15 reference stars with no significant
proper motion in the SSS 
database.

\begin{table}
\caption{Astrometry and photometry of SSSPM\,1930}              
\label{tab_astm1930}      
\centering                                      
\begin{tabular}{l l c c}          
\hline\hline                        
$\alpha, \delta$ (J2000.0) [h~m~s~~\degr~\arcmin~\arcsec]& Epoch & Magnitude & Data \\    
\hline                                   
19 29 41.019  $-$43 10 17.73  & 1978.675 & $B_J=21.077$ & SSS-UK \\
19 29 41.036  $-$43 10 17.96  &          &              & corr. \\
19 29 41.033  $-$43 10 25.42  & 1987.657 & $R=18.315$ & SSS-ESO \\
19 29 41.014  $-$43 10 25.63  &          &            & corr. \\
19 29 40.986  $-$43 10 28.64  & 1991.543 & $I=16.311$ & SSS-UK \\
19 29 41.009  $-$43 10 28.89  &          &            & corr. \\
19 29 40.987  $-$43 10 32.24  & 1995.630 & $R=18.531$ & SSS-UK \\
19 29 41.005  $-$43 10 32.56  &          &            & corr. \\
19 29 41.01~~  $-$43 10 33.0  & 1996.416 & $I=16.261$ & DENIS \\
19 29 40.999  $-$43 10 33.24  &          &            & corr. \\
19 29 40.99~~  $-$43 10 36.8 & 2000.633 & $J=14.794$ & 2MASS \\
                            &          & $H=14.230$ & 2MASS \\
                            &          & $K_s=14.091$ & 2MASS \\
19 29 41.01~~  $-$43 10 37.6  & 2000.503 & $I=16.267$ & DENIS \\
19 29 41.003  $-$43 10 37.61  & 2001.499 &            & corr. \\
19 29 41.002  $-$43 10 38.83  & 2002.934 &            & 3.6m \\
\hline                                             
\end{tabular}
\end{table}

\begin{table}
\caption{Astrometry and photometry of SSSPM\,2003}              
\label{tab_astm2003}      
\centering                                      
\begin{tabular}{l l c c}          
\hline\hline                        
$\alpha, \delta$ (J2000.0) [h~m~s~~\degr~\arcmin~\arcsec]& Epoch & Magnitude & Data \\    
\hline                                   
20 02 52.096  $-$44 32 47.01  & 1980.392 & $B_J\approx22$& SSS-UK \\
20 02 52.116  $-$44 32 47.21  &          &              & corr. \\
20 02 52.097  $-$44 32 53.57  & 1988.448 & $R=19.094$ & SSS-ESO \\
20 02 52.089  $-$44 32 53.93  &          &            & corr. \\
20 02 52.074  $-$44 32 58.90  & 1994.364 & $I=16.075$ & SSS-UK \\
20 02 52.094  $-$44 32 59.09  &          &            & corr. \\
20 02 52.079  $-$44 32 59.65  & 1995.409 & $R=19.564$ & SSS-UK \\
20 02 52.097  $-$44 32 59.88  &          &            & corr. \\
20 02 52.08~~  $-$44 33 03.6 & 1999.627 & $J=13.528$ & 2MASS \\
                            &          & $H=12.987$ & 2MASS \\
                            &          & $K_s=12.586$ & 2MASS \\
20 02 52.06~~  $-$44 33 03.8  & 1998.701 & $I=15.957$ & DENIS \\
20 02 52.099  $-$44 33 04.34  & 2000.458 &            & corr. \\
20 02 52.054  $-$44 33 06.42  & 2002.932 &            & 3.6m \\
\hline                                             
\end{tabular}
\end{table}

The multi-epoch positions used for the proper motion determinations
as well as the photometry are listed 
in Tables~\ref{tab_astm1930} and \ref{tab_astm2003}.
Note that for SSSPM\,2003 there was no original SSS measurement
on the $B_J$ plate since the object was merged with a background 
star which appeared isolated and without a significant
proper motion on all other SSS plates. We measured 
the position of SSSPM\,2003, overlapping with the background star in 
east-west direction on the $B_J$ plate, by eye.

The proper motion of the two HPM objects as determined from the
original SSS data alone (run \#1 in Table~\ref{tab_pm}) was already
in relatively good agreement so that we started to think about a 
possible connection between them. After including all the additional
epochs (run \#2) we got an even better agreement, but the errors
remained very large. The main reason was that for each object one
DENIS position was off by about an arcsecond. We checked the data
taken from the VizieR Catalogue Service (http://vizier.u-strasbg.fr/),
downloaded the corresponding DENIS images, and found a discrepancy
between the epochs given in the catalogue and in the FITS headers
of the images. We preferred the information from the FITS header
which was also consistent with the DENIS strip number given in
the catalogue. After correcting two DENIS epochs the proper motion
errors in run \#3 decreased significantly.

We tried to further improve the proper motion determination by
correcting all the different epoch astrometry relative to the 2MASS 
system. 
The same 15
reference stars, the 2MASS positions of which were already used for
calibrating the astrometry on the acquisition images,
served for a new measurement of our target positions on the
SSS images and for correcting the DENIS catalogue positions (all lines 
with corrected data in Tables~\ref{tab_astm1930} and \ref{tab_astm2003}). 
The resulting proper motion solutions (run \#4 in Table~\ref{tab_pm}) 
are again much more accurate. We adopt this run as our final solution.
It is remarkable that the final results for SSSPM\,1930 and SSSPM\,2003 
are in 
excellent
agreement within the errors.

%

\begin{table}
\caption{Proper motion $(\mu_{\alpha}\cos{\delta},~\mu_{\delta})$ solutions [mas/yr]}              
\label{tab_pm}      
\begin{tabular}{c c c}          
\hline\hline                        
\# & SSSPM\,1930 & SSSPM\,2003 \\    
\hline                                   
1 & $-24.1 \pm 18.4,~ -853.7 \pm ~~4.8$ & $-32.3 \pm 11.7,~ -882.8 \pm 23.2$ \\
2 & $-11.0 \pm ~~8.1,~ -883.5 \pm 17.9$ & $-19.4 \pm ~~5.4,~ -879.7 \pm 25.8$ \\
3 & $-10.5 \pm ~~8.0,~ -871.8 \pm ~~5.8$ & $-19.3 \pm ~~5.0,~ -859.2 \pm 11.7$ \\
4 & $-16.4 \pm ~~3.1,~ -861.6 \pm ~~2.7$ & $-18.0 \pm ~~7.4,~ -855.2 \pm ~~3.8$ \\
\hline                                             
\end{tabular}

Notes:\\
\#1 -- Based on the original 4 SSS epochs for SSSPM\,1930 and 3 SSS epochs (without $B_J$) for SSSPM\,2003 \\
\#2 -- With all the uncorrected positions from Table~\ref{tab_astm1930} (8 epochs) and Table~\ref{tab_astm2003}
       (7 epochs), respectively \\
\#3 -- After correcting two DENIS epochs \\
{\bf \#4 -- Finally adopted run:} with corrected DENIS epochs and after transforming all SSS and DENIS positions to the 2MASS system \\

\end{table}


\section{Probability of a chance alignment of the CPM pair}
\label{sect_chance}

   \begin{figure}
   \centering
   \includegraphics[angle=-90,width=90mm]{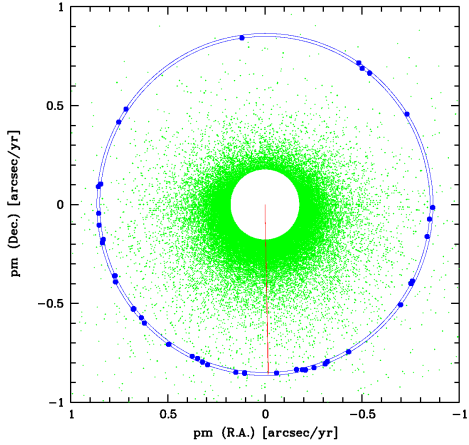}
   \caption{Distribution of NLTT proper motions. Dots show the NLTT
stars with $\mu$$>$0.18~arcsec/yr and proper motion components of less than
$\pm$1~arcsec/yr. The two large circles show proper motion limits of 0.850 and 
0.865~arcsec/yr. Small filled circles mark the NLTT stars falling between
these proper motion limits. The straight line shows the position angles
of the proper motions of SSSPM\,1930/SSSPM\,2003.
               }
              \label{pmd}%
    \end{figure}
%

   \begin{figure}
   \centering
   \includegraphics[angle=-90,width=90mm]{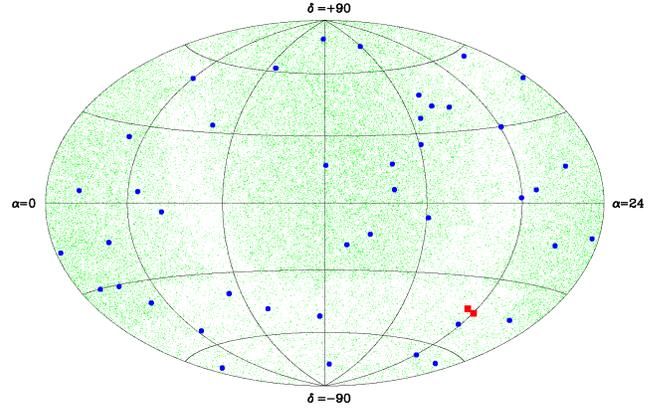}
   \caption{Distribution of all NLTT stars with $\mu$$>$0.18~arcsec/yr
on the sky. Small filled circles mark the NLTT stars with proper motions
between 0.850 and 0.865~arcsec/yr. The two filled squares mark the
positions of SSSPM\,1930 and SSSPM\,2003.
               }
              \label{adp}%
    \end{figure}

Chance alignments of CPM stars have been recently investigated by
L{\'e}pine \& Bongiorno~(\cite{lepine07a}) in a search for new
distant companions to nearby stars. They looked for faint wide CPM
companions to Hipparcos (ESA~\cite{esa97}) stars in the new L{\'e}pine-Shara
Proper Motion-North catalogue (LSPM-North; 
L{\'e}pine \& Shara~\cite{lepine05}). Studying the proper motion
distribution in the LSPM-North and simulating chance alignments between
the two catalogues, they found that real CPM doubles satisfy the condition
$\Delta\theta\Delta\mu<(\mu/0.15)^{3.8}$, 
where $\Delta\theta$ is the angular separation (in arcsec), $\Delta\mu$ is 
the magnitude of the difference between 
the proper motion vectors (in arcsec/yr), and $\mu$ is the mean total proper 
motion of the pair (in arcsec/yr). The range of separations they 
used in their search ($\Delta\theta$$<$1500~arcsec) is much smaller than
in our case ($\Delta\theta$$\approx$21600~arcsec). 
However, the large mean proper motion
($\mu$=0.86~arcsec/yr) and the very small proper motion
difference ($\Delta\mu$$<$0.015~arcsec/yr) of SSSPM\,1930 and SSSPM\,2003 
compensate for the wide separation so that the above condition is fulfilled.

In order to compute the probability of a chance alignment of our CPM 
pair we have compared its proper motions and distribution on the sky with 
known HPM stars from the New Luyten Two Tenths catalogue (NLTT; 
Luyten~\cite{luyten79b}). Fig.~\ref{pmd} shows the central part of the NLTT
proper motion diagram. There are only 46 NLTT stars falling between an
upper limit of 0.865~arcsec/yr and a lower limit of 0.850~arcsecs/yr in the 
total proper motion (i.e. in the region between the two large circles). 
The proper motion limits were selected taking into account the proper
motion errors of the two components of our pair. These are conservative
limits, since their total proper motion agree in fact within 0.007~arcsec/yr.
Their position angles agree within about 0\fdg1 so that there is only
one line seen in Fig.~\ref{pmd} for the position angles of the two
proper motions.

Fig.~\ref{adp} shows the distribution of all NLTT stars 
with $\mu$$>$0.18~arcsec/yr on the sky, with the 46 stars selected from
the previous proper motion diagram marked again by small filled circles. 
Whereas the incompleteness of the NLTT in the Galactic plane and with
$\delta$$<$$-$30\degr\ is clearly seen, the selected stars with relatively
large proper motions seem to be randomly distributed.
With the given density of 46 stars over the 41253 square degrees of the full
sky we compute the probability that two of these stars are separated by
only 6\degr\ as 
46/(41253/36$\pi$)=0.126,
which does not yet take into account the position 
angles of the proper motions. The probability that two of the 46 stars located
in the 
360\degr\,
ring on Fig.~\ref{pmd} have a very small (0\fdg1) difference in
the proper motion orientation is only 
46/(360/0.1)=0.0128,
which does however not take
into account the asymmetrical distribution of the proper motions in 
declination. Using only 37 stars with the corresponding negative proper
motion components 
(half ring of 180\degr), 
we derive a 
somewhat higher
probability of 
37/(180/0.1)=0.0206. 
The 
final probability of a chance alignment in both proper motions and position
angles
can then be derived as 0.126$\times$0.0206=0.0026,
 which is small enough to rule out a random pairing of 
SSSPM\,1930 and SSSPM\,2003.


\section{Spectroscopic distance estimates}
\label{sect_spec}

   \begin{figure}
   \centering
   \includegraphics[angle=-90,width=90mm]{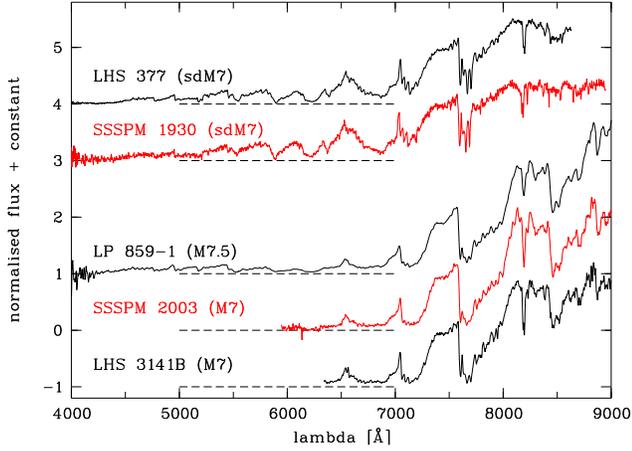}
   \caption{Low-resolution classification spectra of SSSPM\,1930,
SSSPM\,2003 and some comparison objects. The ESO~NTT/EMMI spectrum of 
SSSPM\,1930 is
from Scholz et al.~(\cite{scholz04}), the Keck/LRIS spectrum of LHS\,377 was 
taken from Neill Reid's Web page 
(http://dept.physics.upenn.edu/\~{}inr/ultracool.html). The ESO~3.6m/EFOSC2
spectrum of SSSPM\,2003 is shown together with ESO VLT/FORS1 spectra of two 
comparison objects from Lodieu et al.~(\cite{lodieu05}): LP\,859-1 
and LHS\,3141B. All spectra are flux-calibrated and
normalised at 7500\AA{}.
               }
              \label{5spec}%
    \end{figure}

Spectroscopic follow-up observations of both components of the extremely
wide CPM pair were carried out with the ESO~3.6m
telescope at La Silla between December 5th and 8th, 2002 using the 
EFOSC2 instrument. Based on their red colour, both HPM objects were 
initially treated by us as independent candidates
for previously unknown ultracool (M- and L-type) stars and brown dwarfs
in the immediate Solar neighbourhood. The observations and data reductions 
have already been described in Lodieu et al.~(\cite{lodieu05}). 
For details of the applied spectroscopic classification scheme we 
refer the reader to that paper. 

As already mentioned by Lodieu et al.~(\cite{lodieu05}), the
observing conditions during that particular run were not ideal. Therefore,
only a rather noisy spectrum was obtained for SSSPM\,1930. Despite the low
signal-to-noise ratio of that spectrum (not shown here) we were able to
establish a late subdwarf status with a preliminary spectral type of
sdM5.5. A much higher-quality spectrum
of SSSPM\,1930 was observed by Scholz et al.~(\cite{scholz04}) with the
ESO~NTT equipped with the EMMI instrument (see Fig.~\ref{5spec}).
It is shown together with the spectrum of LHS\,377, the latest-type (sdM7)
object found by Gizis~(\cite{gizis97}) according to his classification
scheme for late-K and M subdwarfs. Based on the spectral indices
used in this scheme, Scholz et al.~(\cite{scholz04}) found also an sdM7
spectral type for SSSPM\,1930. 

The second component of the pair, SSSPM\,2003, turned out to be a normal
late-type M dwarf according to our ESO~3.6m/EFOSC2 spectrum (Fig.~\ref{5spec}).
However, the spectral indices measured for SSSPM\,2003 lead to rather 
different spectral types of M6.1 (from TiO5=0.195 and VOa=2.129) and 
M7.5 (from PC3=1.725; indices as used by
Lodieu et al.~\cite{lodieu05}).
About one third of the M6-M8 dwarfs in Lodieu et al.~(\cite{lodieu05}) do
also show systematically later spectral types if defined by the PC3 index
(see their Fig.~8). The two comparison objects LHS\,3141B (M7) and LP\,859-1 
(M7.5), also shown in Fig.~\ref{5spec}, did not have these classification 
problems in Lodieu et al.~(\cite{lodieu05}). The latter was also found to be an 
M7.5 dwarf by Gizis et al.~(\cite{gizis00}) and Reid et al.~(\cite{reid02}). 
We adopt a spectral type of M7 for SSSPM\,2003 with the
relatively large uncertainty of 1.0
subtypes, taking into account the different typing by spectral indices as 
well as the visual comparison with template spectra.

With the above classification, and using the absolute $J$ magnitudes 
of M6-M8 dwarfs from Scholz, Meusinger \& Jahrei{\ss} (\cite{scholz05}),
we estimate the distance of SSSPM\,2003 as $38^{+10}_{-7}$~pc.  
On the other hand, we use the 2MASS $JHK_S$ magnitudes of LHS\,377 and its
trigonometric parallax listed in Gizis (\cite{gizis97}) for deriving
a distance of $72^{+21}_{-16}$~pc for SSSPM\,1930, conservatively 
assuming an uncertainty 
of 0.5 in the absolute magnitudes of these poorly investigated  
ultracool (sdM7) subdwarfs.
If we further assume our CPM objects, the large proper motions of which
agree to within 1\%, to be at approximately the same distance from the
Sun, we may expect a common distance of roughly $50^{+25}_{-15}$~pc. In
the following we discuss this common distance assumption with respect 
to an expected common space velocity.


\section{Discussion - halo stream and metallicity}
\label{sect_halo}

The assumed distance interval leads to projected physical separations
between 3.5 and 8~pc so that we can rule out a wide binary. Therefore,
we investigate the kinematic connection between the two objects as
possible members in a stellar stream.

The proper motion error of less than 10~mas/yr corresponds with
the assumed distance of 50~pc
to an error in the tangential velocity of less than 2~km/s.
However, the large uncertainty in the distance to our pair
translates with a proper motion of 860~mas/yr
to a wide tangential velocity interval of $204^{+102}_{-61}$~km/s.
A radial velocity (RV) measurement is available for only one of
the components, SSSPM\,1930, 
with $-262\pm25$~km/s from Scholz et al. (\cite{scholz04}).
Therefore, the errors in the $UVW$ space velocity components are
dominated by the uncertain distance estimates and by the (partly lacking)
RV measurements, whereas the proper motion errors play only a minor role.

   \begin{figure}
   \centering
   \includegraphics[angle=-90,width=130mm]{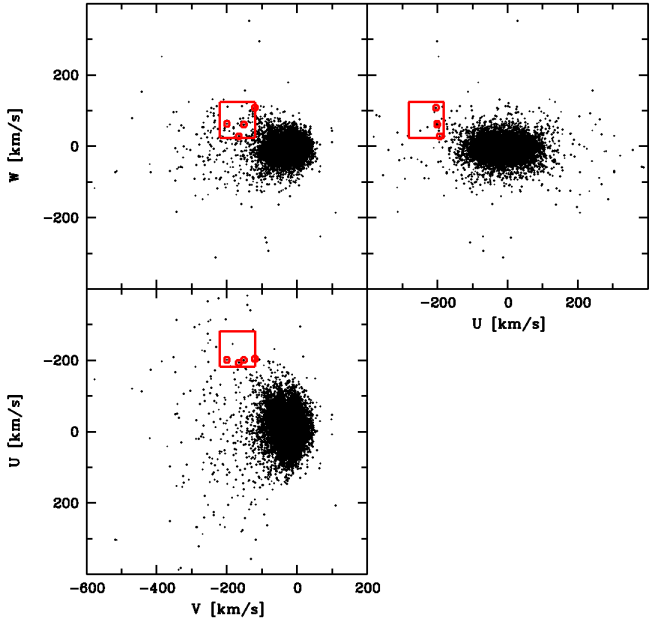}
   \caption{$UVW$ velocities of 35682 CRVAD-2 stars (dots). Four stars
            with velocity components within 50~km/s (squares) of the mean
            velocity of SSSPM\,1930 and SSSPM\,2003 are marked (circles).
               }
              \label{uvw}%
    \end{figure}

We have computed possible heliocentric
Galactic space velocities $UVW$ (Johnson \& Soderblom \cite{johnson87}),
positive towards the Galactic Centre, Galactic rotation,
and North Galactic Pole,
separately for the two objects using their positions, 
the individual proper motions (Table~\ref{tab_pm}, run \#4), and a wide
parameter space for the parallaxes (from 10 to 40~mas in steps of 0.5~mas)
and RVs (from $-$300 to $+$300~km/s in steps of 5~km/s).
An exact agreement, to within 5~km/s for all three $UVW$ velocity components,
could only be achieved for positive RVs, where with distances 
of the order of 50~pc the RVs were required to be larger than $+$140~km/s.
Generally, the agreement in the space velocities was achieved with slightly
larger parallaxes ($+$1~mas) and RVs ($+$5~km/s) for SSSPM\,2003 compared
to those of SSSPM\,1930.

However, if we consider the large negative RV of SSSPM\,1930 measured by 
Scholz et al. (\cite{scholz04}) to be approximately correct, 
then we can only achieve
an agreement in all three space velocity components to within 25~km/s.
In this case, SSSPM\,2003 would lie at $\approx$32~pc but SSSPM\,1930 
would be much too close ($<$27~pc) compared to our 
spectroscopic estimate. If we allow for differences in the $UVW$ components
of up to 30~km/s, then we get the following solution with individual
distances close to our preferred common distance of about 50~pc:\\

SSSPM\,1930: $d$=44~pc, RV=$-$235~km/s \\
$(U,V,W) = (-245 \pm  25, -157 \pm  51,  +59 \pm  16)$~km/s\\

SSSPM\,2003: $d$=50~pc, RV=$-$220~km/s \\
$(U,V,W) = (-219 \pm  24, -183 \pm  66,  +89 \pm  15)$~km/s\\

Opposite to our spectroscopic distances, SSSPM\,2003 turns
out to be more distant and having a smaller negative RV value than
SSSPM\,1930
in all solutions with the condition of similar $UVW$ components.
An unresolved equal-mass binary status of SSSPM\,2003 would
lead to a corrected spectroscopic distance of $\approx$54~pc
and help to explain the above mentioned discrepancy.

Although we are aware of the uncertainties in our input data (distances
and RVs) and in the assumption of a common space motion of the pair,
we can further speculate on the consequences of the computed $UVW$.
According to Chiba \& Beers (\cite{chiba00}) the above listed large $U$ and $V$
and moderately large $W$ values are more typical of Galactic halo rather than
of thick disk stars. To our knowledge, no halo stream with comparable
velocity components has been identified so far.
The Catalogue of Radial Velocities with Astrometric Data
(CRVAD-2; Kharchenko et al.~\cite{kharchenko07}) lists the RVs of about 
55\,000 bright stars together with their proper motions and trigonometric
parallaxes if available. We have computed $UVW$ for all CRVAD-2 stars  
whose parallaxes are more than three times larger than their errors (35682 stars) 
and searched 
for stars with velocity components similar (within 50~km/s) to the mean 
$UVW$ of our wide pair (Fig.~\ref{uvw}). Only four  
stars (marked on Fig.~\ref{uvw}) were found to satisfy
these criteria: Hip 10449, 12294, 18742, 81528. Two of them are in the
equatorial zone, the others have large declinations with different signs.
The nearest (Hip 10449) is at $\approx$60~pc, the others lie between 150~pc
and 420~pc.
Three of these F8- to K0-type stars are widely discussed in the 
literature as thick disk/halo stars. However, their metallicities are 
different (-2.5 $<$ [Fe/H] $<$ -0.7 
dex,
according to the Simbad database).


The similarity of the two subdwarfs LHS\,377 and SSSPM\,1930 and
their clearly different spectra compared to the normal dwarfs (including
SSSPM\,2003) is well seen in Fig.~\ref{5spec}. However, if we apply the 
procedure proposed by L{\'e}pine et al.~(\cite{lepine07}) to derive
revised metallicity classes from the measured spectral indices 
CaH2, CaH3 and TiO5 of LHS\,377 (taken from Gizis \cite{gizis97}) and SSSPM\,1930 
(from Scholz et al. \cite{scholz04}) we get a $\zeta_{\text{TiO/CaH}}$
of 0.994 and 0.918, respectively. Consequently, the two objects would
be classified as normal dwarfs (dM) instead of subdwarfs (sdM), M7.2 and
M6.9, respectively. 
Unfortunately, their indices fall exactly in the nonvalidated region 
(0.2$<$TiO5$<$0.7, 0.5$<$CaH2+CaH3$<$0.9) of the classification 
diagram used by L{\'e}pine et al.~(\cite{lepine07}), where they
mentioned a different slope of the iso-metallicity lines obtained
from model spectra. Therefore, we prefer our sdM7 classification 
of SSSPM\,1930, hinting at a lower metallicity compared to the M7 
dwarf SSSPM\,2003, until better measurements and classification 
schemes will be available. The metallicity problem does not affect
our crude estimates for the distance and space velocity of the pair.

The two objects described in this paper form the widest-known pair 
of (very low-mass) stars with a large proper motion in common.
More detailed investigations of this extreme pair
are certainly needed, in particular accurate RV measurements
of both components. Trigonometric parallaxes will further improve the
velocity components and help to explain the physical association of the 
objects with each other and their possible membership in a Galactic 
halo stream.


\begin{acknowledgements}

   We would like to thank the anonymous referee for his/her comments 
   and suggestions, which helped us to improve the paper.
   We acknowledge the use of
   the Simbad database  and the VizieR Catalogue Service operated at the 
   Centre de Donn\'ees astronomiques de Strasbourg (CDS), France.
   This study was supported by the German DFG grant 436 RUS 113/757/0-2, and
   by the Russian RFBR grant 07-02-91566. 
\end{acknowledgements}

\end{document}